\documentstyle[aps,prd,epsfig]{revtex} 
\newcommand{\be}{\begin{equation}} 
\newcommand{\ee}{\end{equation}} 
\newcommand{\bea}{\begin{eqnarray}} 
\newcommand{\eea}{\end{eqnarray}} 
\newcommand{\bdm}{\begin{displaymath}} 
\newcommand{\edm}{\end{displaymath}}

\newcommand{\we}{\wedge} 

\newcommand{\CC}{\mbox{$I \! \! \! \! C$}} 
\newcommand{\RR}{\mbox{$I \! \! R$}}
\newcommand{\gtens}{\mbox{\boldmath $g$}} 
\newcommand{\bmx}{\mbox{\boldmath $x$}} 
\newcommand{\bmy}{\mbox{\boldmath $y$}} 
\newcommand{\bmxi}{\mbox{\boldmath $\xi$}} 
\newcommand{\bme}{\mbox{\boldmath $e$}} 
\newcommand{\bmal}{\mbox{\boldmath $\alpha$}} 
\newcommand{\bmnab}{\mbox{\boldmath $\nabla$}} 
\newcommand{\bmeta}{\mbox{\boldmath $\eta$}} 
\newcommand{\bmgam}{\mbox{\boldmath $\gamma$}} 
\newcommand{\zbar}{\bar{z}} 
\newcommand{\wiz}{\partial_{z}} 
\newcommand{\wizb}{\partial_{\bar{z}}} 
\begin{document}
\title{Complex Formulation of \\ Lensing Theory and Applications} 
\author{Norbert Straumann}
\address{Institute for Theoretical Physics \\
The University of Zurich \\
CH--8057 Zurich, Switzerland}
\date{\today}
\maketitle
\begin{abstract}
The elegance and usefulness of a complex formulation of 
the basic lensing equations is demonstrated with a 
number of applications. Using standard tools of complex 
function theory, we present, for instance, a new proof of the 
fact that the number of images produced by a 
regular lens is always odd, provided that the source is not 
located on a caustic. Several differential and integral 
relations between the mean curvature and the (reduced) shear 
are also derived. These emerge almost automatically from 
complex differentiations of the differential of the lens map, 
together with Stokes' theorem for complex valued $1$-forms.
\end{abstract}
\pacs{....}
\section{Introduction}
Gravitational lensing has become one of the most important fields 
in present day astro\-nomy. The enormous activity in 
this area has largely been driven by considerable improvements of 
observational capabilities. Gravitational lensing has the distinguished 
feature of being independent of the nature and the physical state of the 
deflecting mass. It is therefore perfectly suited to study the dark 
matter in the Universe \cite{1}, \cite{2}. 

One of the issues which has recently attracted a lot of attention is 
concerned with parameter-free reconstructions of projected mass 
distributions from weak lensing data. (For a recent review, see \cite{3}.) 
Thanks to new wide-field cameras and imaging with $8m$-class 
telescopes, the quality of the data is expected to increase rapidly. Initiated 
by a paper of Kaiser and Squires \cite{4}, a considerable amount of 
theoretical work on various reconstruction methods has recently also been 
carried out \cite{5}, \cite{6}. The main problem consists in the task to make 
optimal use of limited noisy data in a parameter-free manner, that is, 
without modeling the lens.

In the present paper we take up some of the theoretical discussions and 
demonstrate rather systematically that the complex formulation of lensing 
theory often simplifies
things considerably. In particular, a number of equations which are used 
in mass reconstructions, emerge almost automatically.

In outline, the paper is organized as follows:
For reasons of self-consistency, we provide in Section 2 a brief 
derivation of the basic lensing equations that are used
in the remainder of the paper. These are then translated in Section 
3 into a complex formulation, where some mathematical tools are 
recapitulated as well. It will turn out that the reconstruction 
problem is basically equivalent to the task of solving the so-called 
Beltrami equation,
at least for noncritical lenses. This part of the paper has 
considerable overlap with \cite{7}. Turning to applications in 
Section 4, we give -- as far as we know -- a new proof of the fact 
that for a regular lens the number of images is always odd, 
provided that the source is not located on a caustic. The proof uses 
only standard tools of complex 
analysis, which are, for instance, familiar from derivations of the 
theorem of residues. One of these formulas is an explicit expression 
for the index of a closed curve relative to a given point. Next, we 
derive several relations between the mean convergence and the 
(reduced) shear by (repeated) applications of the complex 
differential operators $\partial / \partial z$ and $\partial / \partial 
\bar{z}$ to the differential of the lens map. Several other useful 
relations for lensing reconstructions, involving integrals over 
bounded domains, are derived at the end of the paper.

The purpose of this article is mainly methodo\-logical.
We hope that others will take advantage of it, especially
in teaching the pleasant field of gravitational lensing.

\section{Basic Lensing Equations}

For the benefit of those readers who have not studied the 
extensive monograph of Schneider, Ehlers and Falco \cite{1}, we 
start by giving a brief derivation of the basic lensing equations.

The conceptual basis of gravitational lensing theory is extremely 
simple. This is at the same time one of the main reasons why it is 
so important for the astronomical study of mass distributions on 
all scales. For all practical purposes the ray approximation for 
light propagation is sufficiently exact. In this limit the rays 
correspond to null geodesics in a given gravitational field 
$\gtens$, 
and the evolution of the polarization vector is governed by the law 
of parallel transport. (These laws can be deduced 
from Maxwell's equations \cite{8}.) The null rays are orthogonal to the 
surfaces of constant phase, $\{S = \mbox{const}\}$, where $S$ is 
subject to the eikonal equation
\be
g^{\mu \nu} \, \partial_{\mu} S \, 
\partial_{\nu} S \, = \, 0 \, .
\label{eikonal}
\ee
For sufficiently strong lenses the wave fronts develop edges and 
self-intersections. Clearly, an observer behind such folded fronts 
sees more than a single image. This is the region of what is 
called strong lensing and occurs astronomically only rarely.

Here we restrict ourselves to almost Newtonian, asymptotically 
flat situations. Generalizations to the cosmological context are 
easy and basically amount to interpret all distances in the 
formulas given below
as angular distances. (For details we refer again to \cite{1}, 
hereafter quoted as SEF). The metric is then given by
\be
\gtens \, = \, \left( 1 + 2 \, U \right) \, dt^2 \, - \, \left( 1 - 2 \, U 
\right) \, d\bmx^2 \, , \label{metric-1}
\ee
where $U$ is the Newtonian potential. The spatial part of a light 
ray satisfies Fermat's principle,
\be
\delta \, \int \frac{d \sigma}{\sqrt{g_{00}}} \, = \, 0 \, , 
\label{Fermat}
\ee
for variations with fixed end points \cite{8}.
Here $d \sigma^2$ denotes the spatial part of the metric 
(\ref{metric-1}). 

All this can be summarized by saying that gravitational lensing 
theory is just usual ray optics with the refraction index
\be
n(\bmx) \, = \, 1 \, - \, 2 \, U(\bmx) \, .
\label{index} 
\ee
In particular, the ray equation holds,
\be
\frac{d}{ds} \, \left( n \, \frac{d \bmx}{ds} \right)
\, = \, \bmnab n \, ,
\label{ray-eq}
\ee
where $s$ is the euclidean path length parameter. (Since light 
deflection is a scattering process, we can from now on forget 
about non-euclidean geometry.)

In terms of the unit tangent vector $\bme = d \bmx/ds$, 
eq. (\ref{ray-eq}) can be written in sufficient approximation as
\be
\frac{d}{ds} \, \bme \, = \, -2 \, \bmnab_{\! \perp} U \, , \label{ray-
eq-2}
\ee
where $\bmnab_{\! \perp}$ denotes the transverse derivative, 
$\bmnab_{\! \perp} = \bmnab - (\bme,\bmnab) \bme$.
This gives for the deflection angle 
$\hat{\bmal} = \bme_{in} - \bme_{fin}$,
with initial and final directions
$\bme_{in}$ and $\bme_{fin}$, respectively,
\be
\hat{\bmal} \, = \, 2 \, 
\int_{u.p.} \bmnab_{\! \perp} U \, ds \, ,
\label{angle-1}
\ee
where the integral is taken over the unperturbed path (u.p.). Here, we insert 
the expression for the Newtonian potential of
a mass density $\rho(\bmx)$. In the well-justified approximation 
where the extension of the lens (for instance a cluster of galaxies) 
is much smaller than the distances of the observer and the source to 
the lens, one finds readily (SEF, Chapter 4)
\be
\hat{\bmal}(\bmxi) \, = \, 4 \, G \, \int_{\RR^2} \frac{\bmxi - 
\bmxi'}{\mid \bmxi - \bmxi' \mid^2} \, \Sigma (\bmxi') \, d^2 \xi' \, 
,
\label{angle-2} 
\ee
where $\Sigma(\bmxi)$ denotes the projected mass density 
on the lens plane. (For a point mass this reduces to Einstein's prediction 
of light deflection.)

Combining this with elementary geometry, we arrive at the lens map for 
a given $\Sigma(\bmxi)$. From Fig.\/1, which summarizes the notation 
of SEF, we read off the lens equation \be
\bmeta \, = \, \frac{D_s}{D_d} \, \bmxi \, -\, 
D_{ds} \, \hat{\bmal}(\bmxi) \, ,
\label{lens-eq}
\ee
which defines a map from the lens plane to the source plane. 

It is 
convenient to write this in dimensionless form. Let $\xi_0$ be a length 
parameter in the lens plane (whose choice will depend on the specific 
problem), and let $\eta_0$ be the corresponding scaled length in the 
source plane,
$\eta_0 = (D_s/D_d) \xi_0$. 
We set $\bmx = \bmxi/\xi_0$, $\bmy = \bmeta/\eta_0$
and (following SEF)
\be
\kappa(\bmx) \, = 
\, \frac{\Sigma(\xi_0 \/ \bmx)}{\Sigma_{crit}} \, , \; \; \; \; \bmal(\bmx) 
\, = 
\, \frac{D_d \, D_{ds}}{\xi_0 \/ D_s} \,
\hat{\bmal}(\xi_0 \/ \bmx) \, ,
\label{210}
\ee
with
\bdm
\Sigma_{crit} \, = \, \frac{c^2}{4 \pi \/ G} \, 
\frac{D_s}{D_d \/ D_{ds}} \, .
\edm
Then eq. (\ref{lens-eq}) reads as follows
\be
\bmy \, = \, \bmx \, - \, \bmal(\bmx) \, , \label{11}
\ee
whereby eq. (\ref{angle-2}) translates to
\be
\bmal(\bmx) \, = \, \frac{1}{\pi} \, \int_{\RR^2} 
\frac{\bmx - \bmx'}{\mid \bmx - \bmx' \mid^2} \, 
\kappa (\bmx') \, d^2 x' \, .
\label{angle-3} 
\ee
It is obvious that $\bmal$ is a gradient of a two-dimensional 
Newtonian potential:
\begin{center}
\begin{figure}[here]
\epsfig{file=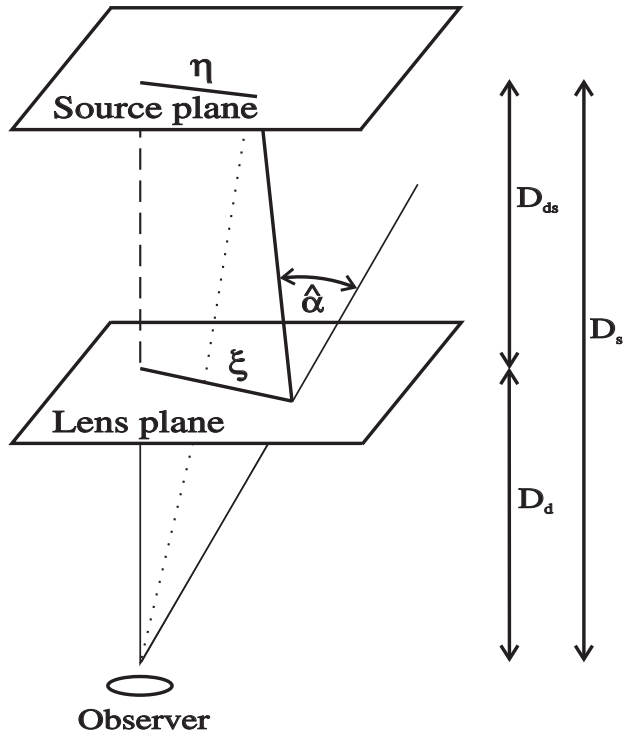,width=8.8cm}
\caption{Notation adopted for the describtion of the lens geometry.} 
\label{geometry.eps}
\end{figure}
\end{center}
\be
\bmal \, = \, \bmnab \psi \, , \; \; \; \; 
\psi \, = \, 2 \/ G \ast \kappa \, , \; \; \; \; 
G(\bmx) \, = \, \frac{1}{2 \pi} \, \ln \mid \bmx \mid \,\,\,\,\,\,\text{($*$ denotes
convolution)}\, . 
\label{213}
\ee
Since $G$ is a fundamental solution of the two-dimensional Laplace 
operator, $\psi$ satisfies the two-dimensional Poisson equation
\be
\Delta \psi \, = \, 2 \, \kappa \, .
\label{214}
\ee

For the differential $D \varphi$ of the map $\varphi\/:\/\RR^2 
\rightarrow \RR^2$, defined by 
eq. (\ref{11}), we use the standard parametrization \be
D \varphi \, = \,
\left( \begin{array}{cc}
1 - \kappa - \gamma_1 & - \gamma_2 \\
- \gamma_2 & 1 - \kappa + \gamma_1
\end{array} \right) \, ,
\label{215}
\ee
in terms of the mean (Ricci-) curvature $\kappa$, determined by the 
trace of $D \varphi$, and the (Weyl-) shear vector $\bmgam = 
(\gamma_1,\gamma_2)$.
The eigenvalues of the symmetric matrix $D \varphi$ are
$1 - \kappa \mp \mid \bmgam \mid$. The critical  
curves, satisfying $\det \left(D \varphi\right) = 0$, are given by
\be
(1 - \kappa)^2 \, - \, \mid \bmgam \mid^2 \, = \, 0 \, . \label{crit-
curves}
\ee
The caustics are the images of these critical curves.
In the vicinity of a caustic the amplification $\mu$, given by
\be
\mu \, = \, \frac{1}{\mid \mbox{det}(D \varphi) \mid} \, , 
\label{217}
\ee
becomes very large.

In passing, we note that the lens map (\ref{11}) can also 
be written as
\be
\bmnab_{{\tiny\bmx}} \phi \, = \, 0 \, , \; \; \;
\mbox{with} \; \;
\phi(\bmx,\bmy) \, = \, \frac{1}{2}(\bmx-\bmy)^2 \, - 
\, \psi(\bmx) \, . \label{lens-eq-4}
\ee
This reflects the Fermat principle. Indeed, the 
delay of arrival times is directly given by the Fermat 
potential $\phi$:
\be
\triangle t \, = \, \xi_0^2 \, \frac{D_s}{D_d\/D_{ds}} \, 
\phi(\bmx,\bmy) \, .
\label{time-delay}
\ee

Examples of various lens maps are discussed extensively in 
Chapter 8 of SEF. Two standard cases are (with suitable 
choices of $\xi_0$):
\bea
& & \mbox{Schwarzschild lens:} \; \;
\bmy \, = \, \bmx \, - \bmx / \! \mid \bmx \mid^2 \, ; 
\label{220}\\
& & \mbox{singular isothermal lens:} \; \; 
\bmy \, = \, \bmx \, - \bmx / \! \mid \bmx \mid \, . \label{221}
\eea

It is worth recalling the following general fact:
In 1955, in a pioneering work of modern singularity theory, H. 
Whitney \cite{9} studied generic properties of smooth mappings of 
the plane into itself and proved that the subset of mappings which 
have only fold and cusp singularities contains an open and dense set 
(with respect to the Whitney topology). Moreover, those maps of 
this set which satisfy a few mild global conditions are also stable. 
Clearly, these results are highly relevant to gravitational lensing. 
For realistic lenses we will only have folds and cusps, and no 
singularities of higher order.

\section{Complex Formulation}

In this section we translate the basic lensing equations into a 
complex formulation. It will turn out that this is not only elegant, 
but also quite useful, because one
can then apply various tools and techniques of complex analysis. 
This has also been noted before by other authors \cite{7}.
\subsection{Mathematical Preliminaries}
We use standard notation when identifying $\RR^2$
with $\CC$, by writing $z = x + i y$ for $(x,y) \in \RR^2$ and 
$dz = dx + i\/dy$, $d\zbar = dx - i\/dy$ for the corresponding 
basis of $1$-forms. In terms of the Wirtinger derivatives,
\be
\wiz \, \equiv \frac{\partial}{\partial z} \, = \, 
\frac{1}{2} \left( \frac{\partial}{\partial x} \, - \, i \, 
\frac{\partial}{\partial y} \right) \, , \; \; \; \; \wizb \, \equiv 
\frac{\partial}{\partial \zbar} \, = \, \frac{1}{2} \left( 
\frac{\partial}{\partial x} \, + \, i \, \frac{\partial}{\partial y} 
\right) \, , 
\label{Wirt} 
\ee
the differential of any smooth complex function $f$ on $\CC$ has 
the representation
\be
d\/f \, = \, \frac{\partial f}{\partial z} \, dz \, + \,  \frac{\partial 
f}{\partial \zbar} \, d \zbar \, . \label{diff-f}
\ee
We shall also write $f_z$ and $f_{\zbar}$ for $\wiz f$ 
and $\wizb f$, respectively.
A function $f$ is holomorphic if and only if $\wizb f = 
0$. In terms of the Wirtinger derivatives the Laplacian is 
given by
\be
\Delta \, = \, 4 \, \wiz \, \wizb \, . \label{laplace}
\ee

We shall make repeated use of Stokes' theorem
for complex-valued differential forms on $\CC$ 
(or an open subset): If $\Omega$ is a compact 
subset of $\,\CC$ with smooth boundary $\partial \Omega$, 
then for every complex differential $1$-form $\omega$ \be
\int_{\Omega} d \omega \, = \, 
\int_{\partial \Omega} \omega \, .
\label{34}
\ee
An immediate corollary of eq. (\ref{34})
is the Cauchy-Green formula: For a smooth function
$f$ we consider
\be
\omega \, = \, f \, \frac{dz}{z - \zeta} \, , \label{def-om}
\ee
and apply Stokes' theorem (\ref{34}) for
$\Omega$ minus an $\varepsilon$-disk with center $\zeta$. In 
the limit $\varepsilon \rightarrow 0$ we obtain
\be
f(\zeta) \, = \, \frac{1}{2 \pi\/i} \, \int_{\partial \Omega} 
\frac{f(z)}{z - \zeta} \, dz \, + \, 
\frac{1}{2 \pi\/i} \,
\int_{\Omega} \frac{f_{\zbar}(z)}{z - \zeta} 
\, dz \we d \zbar \, .
\label{C-G-F}
\ee
For holomorphic functions the second integral is absent. (Note 
that $dz \we d \zbar = -2i dx \we dy$.)

The {\em dilatation\/} or {\em Beltrami coefficient\/} $\nu = 
\nu_f$ of a smooth function $f$ is defined by \be
f_{\zbar} \, = \, \nu_f \, f_z \, ,
\label{def-Bel} 
\ee
and this equation is also called
{\em Beltrami equation\/}. Since the Jacobian $J_f$ of $f$ is 
given by 
\be
J_f \, = \, \mid f_z \mid^{\,2} \, - \,  
\mid f_{\zbar} \mid^{\,2} \, ,
\label{38}
\ee
we conclude that $\mid \! \nu_f \! \mid < 1$ if $f$ preserves 
orientation and $\nu_f = 0$ if and only if $f$ is conformal. 
For the interpretation of $\nu_f$ we consider 
the infinitesimal ellipse field by assigning to each $z\in \CC$ the ellipse that is mapped to a circle 
by $f$. 
As indicated in Fig.~2, the argument of the major axis of this infinitesimal ellipse is $\left\lbrack 
\pi+arg(\nu_f)\right\rbrack/2$, and the eccentricity $\epsilon$ is 
\be
\epsilon = \frac{\vert f_z\vert-\vert f_{\zbar}\vert}{\vert f_z\vert+\vert f_{\zbar}\vert} = \frac{1-
\vert \nu_f \vert}{1+\vert \nu_f \vert}\,.
\ee\label{39}
\begin{center}
\begin{figure}
\epsfig{file=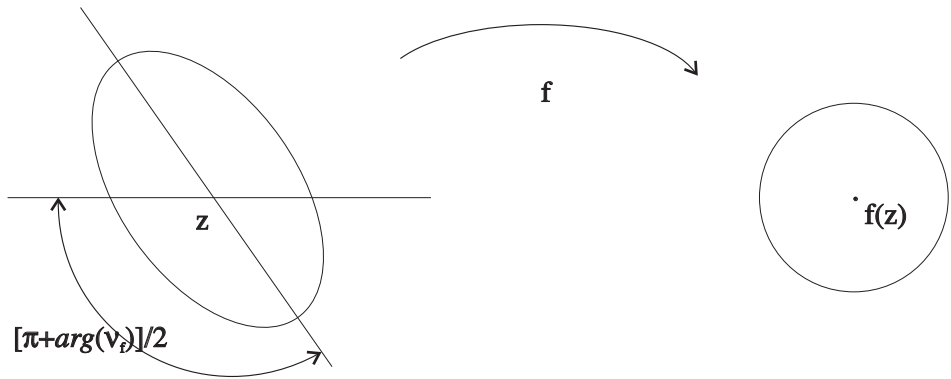,width=12.8cm}
\caption{Geometrical interpretation of the Beltrami parameter.} 
\label{lens.eps}
\end{figure}
\end{center}
Solving the Beltrami equation (\ref{def-Bel})  is then equivalent to finding a function $f$ whose associated 
ellipse field coincides with a prescribed $\nu$. We shall see
that this is just the inversion problem in gravitational lensing. Weak gravitational
lensing corresponds to quasiconformal maps. A smooth map $f$ is
{\em k-conformal\/} if its Beltrami parameter $\nu_f$ satisfies $\vert\nu_f\vert\leq k < 1$. This means 
geometrically that there is a fixed bound on the stretching of $f$ in any given direction compared to any other 
direction.

We now quote an existence and uniqueness theorem for the Beltrami equation. For a fixed $k$ with $0<k<1$ let 
$L^{\infty}(k,\,R)$ denote the measurable functions on $\CC$ bounded by 
$k$ and supported in $\left\lbrace z\in\CC\vert\,\vert z \vert < R\right\rbrace$.

{ \it \underline{Theorem}:
For $\nu \in L^{\infty}(z,\,R)$, there is a complex function $f$ on $\CC$, normalized so that $f(z)=z+{\cal O}(1/z)$ 
at $\infty$, with distributional derivatives satisfying the Beltrami equation $f_{\zbar} \, = \, \nu \, f_z $, and such 
that $f_{\zbar} $ and $f_{z}-1 $ belong to $L^p$ for a $p>2$ sufficiently close to 2. Any such $f$ is unique. The 
solution $f$ is a homeomorphism of $\CC$, which is holomorphic on any open set on which $\nu = 0$. If $\nu \in C^1$ 
and $\nu_z \in C^1$, then $f\in C^1$.}
 
A proof of this theorem can, for instance, be found in \cite{10}.

The reconstruction problem (for noncritical lensing) will lead to the inhomogeneous Cauchy-Riemann equation
\begin{equation}\label{310}
\partial_{\zbar}f=h\,.
\end{equation}
In case the smooth function $h$ has compact support, the Cauchy-Green formula (\ref{C-G-F})
provides one solution:
\begin{equation}\label{311}
f\left(\zeta\right) = \frac{1}{2\pi i}\int_{\,\,\CC}\frac{h\left( z\right)}{z-\zeta}dz\wedge d\zbar\,.
\end{equation}
Obviously, $f$ is only determined up to an additive holomorphic function. If the solution is assumed to be bounded, 
$f$ is {\em unique up to an additive constant\/}.

From the solution (\protect\ref{311}) we see that $(\pi z)^{-1}$ is a fundamental solution of the differential operator 
$\partial_{\zbar}$,
\be
\frac{1}{\pi}\partial_{\zbar}\left(\frac{1}{z}\right) = \delta\,,
\ee\label{312}
because  (\ref{311}) can be written as 
\be
f = \frac{1}{\pi}\frac{1}{z}*h\,.
\ee\label{313}

A special case of the so-called Dolbaut Lemma in several complex variables implies that  one may drop the 
assumption that $h$ has compact support:

{ \it \underline{Theorem}:
For any smooth function $h$ on $\CC$ there exists a smooth function $f$ such that (\ref{310}) holds.}

For a complete proof, see Chapter 2 of \cite{11}.

As an easy consequence we have the 

{ \it \underline{Corollary}:
For any smooth function $h$ there exists a smooth solution of the Poisson equation
$\Delta f=h$.}

In the following we often use the abreviations 
$\partial\equiv\partial_z$, $\bar\partial\equiv\partial_{\zbar}$.

\subsection{The complex Lens Mapping and its Differential}

The lens mapping $\varphi:\RR^2\longmapsto\RR^2$,
\begin{equation}\label{314}
\bmy=\varphi\left(\bmx\right) = \bmx-\bmnab\psi\left(\bmx\right)\,,
\end{equation}
is now written as $f: \CC\longmapsto\CC$, $w=f(z)$ with $z=x_{1}+ix_{2}$, 
$w=y_{1}+iy_{2}$. We have 
\begin{equation}\label{315}
f\left( z\right) = z-2\bar\partial\psi
\end{equation}
or
\begin{equation}\label{316}
f=\bar\partial\left(z\zbar-2\psi\right)\,.
\end{equation}
Eq. (\protect\ref{214}) becomes 
\begin{equation}\label{317}
2\partial\bar\partial\psi = \kappa\,.
\end{equation}
The differential of $f$ will be very important. From (\protect\ref{315}) and 
(\protect\ref{317}) we obtain
$$
df=(1-\kappa) dz-2\bar\partial^2\psi d\zbar\,.
$$
But
$$
\bar\partial^2\psi = \frac{1}{4}\left(\partial_1^2-\partial_2^2\right)\psi+\frac{i}{2}
\partial_1\partial_2\psi = \frac{1}{2}\left(\gamma_1+i\gamma_2\right),
$$
according to the original definition (\protect\ref{215}) of the shear vector. Introducing the 
complex shear
\begin{equation}\label{318}
\gamma = \gamma_{1}+i\gamma_{2}
\end{equation}
we obtain 
\begin{equation}\label{319}
df=(1-\kappa)dz-\gamma d\zbar\,.
\end{equation}
Hence, the Beltrami parameter $\nu_f$ of the lens map is given by
\begin{equation}\label{320}
\nu_f = \frac{\gamma}{1-\kappa}\,.
\end{equation}
This agrees with the reduced shear introduced by Schneider and Seitz
\cite{12}. 

The examples 
(\protect\ref{220}) and (\protect\ref{221}) become:
\begin{eqnarray}
& & \text{Schwarzschild lens:}\,f(z)=z-\frac{1}{\zbar},\,\nu_f = \frac{1}{\vert
z\vert^2}\,;\label{321} \\
& & \text{singular isothermal lens:}\,f(z)=z-\frac{z}{\vert\zbar\vert},\,\nu_f = 
\frac{z}{\zbar\vert z\vert}\,.\label{322}
\end{eqnarray}

For reference, we note that the amplification $\mu$ is according to (\protect\ref{217}), 
(\protect\ref{38}) and (\protect\ref{319}) given by
\begin{equation}\label{323}
\mu^{-1} = \vert J_f\vert = \vert\vert\partial f\vert^2-\vert\bar\partial f\vert^2\vert = \vert 
(1-\kappa)^2-\vert \gamma\vert^2\vert\,.
\end{equation}

\section{Applications}

The usefulness of the complex formulation will be illustrated in this section with several 
applications. No new results are obtained, but some of the derivations become simpler and 
more natural.

\subsection{Number of Images for a regular Lens}

The important fact that the number of images for a regular lens is always odd, provided the 
source does not lie on a caustic, is traditionally proven with the help of some elements of 
Morse theory \cite{1}. We now give a proof which uses only standard tools of complex 
function theory that are used, for example, in the derivation of the theorem of residues. In 
particular, we make use of the following analytic formula for the index of a closed 
(rectifiable) curve $\gamma$ relative to a point $a\not\in\gamma$:
\begin{equation}\label{41}
ind_{\gamma}\left( a\right)=\frac{1}{2\pi i}\int_{\gamma}\frac{dz}{z-a}\,.
\end{equation}
This index is equal to the winding number of $\gamma$ around $a$ and hence an integer. 
Furthermore, it is a homotopic invariant, changes sign under orientation reversion, and is 
additive under composition of closed curves (see, e.g., Chapter IV of
\cite{13}).

Consider now a point $w_{\circ}$ in the source plane with images 
$f^{-1}\left(w_{\circ}\right) = \left\lbrace z_1,\dots,\,z_N\right\rbrace$ in the lens plane. 
The complex 1-form
\begin{equation}\label{42}
\omega = \frac{1}{2\pi i}\frac{df}{f-w_{\circ}}
\end{equation}
is regular on $\CC\backslash\bigcup_{j}D_{\epsilon}\left( z_{j}\right)$, where 
$D_{\epsilon}\left( a\right)$ denotes the closed disk with center $a$ and radius $\epsilon$. It 
is also closed, and therefore Stokes' theorem (\protect\ref{34}) gives
\begin{equation}\label{43}
\frac{1}{2\pi i}\int_{\partial D_{R}\left( 0\right)}\frac{df}{f-w_{\circ}} = 
\sum_{j=1}^{N}\frac{1}{2\pi i}\int_{\partial D_{\epsilon}\left( z_j\right)}\frac{df}{f-
w_{\circ}}\,.
\end{equation}
Now, for a closed curve $\gamma$ we have by the transformation formula of integrals and 
(\protect\ref{41})
\begin{equation}\label{44}
\frac{1}{2\pi i}\int_{\gamma}\frac{df}{f-w_{\circ}} = \frac{1}{2\pi 
i}\int_{f\circ\gamma}\frac{dw}{w-w_{\circ}} = ind_{f\circ\gamma}\left(w_{\circ}\right)\,.
\end{equation}
Asymptotically the lens map approaches the identity, and hence the left hand side of 
(\protect\ref{43}) is equal to $1$ for $R$ sufficiently large. Therefore, we have
\begin{equation}\label{45}
1=\sum_{j=1}^{N}ind_{f\circ\partial D_{\epsilon}\left( 
z_j\right)}\left(w_{\circ}\right)=n_1-n_{-1}+2\left(n_2-n_{-2}\right)+\dots,
\end{equation}
where $n_{\lambda}$ denotes the number of $z_j$ in $\left\lbrace 
z_1,\dots,\,z_N\right\rbrace$ for which the index in (\protect\ref{45}) is equal to $\lambda$.

For the special case, when $w_{\circ}$ is not on a caustic, the Jacobians $J_{f}(z_j)$ do not 
vanish and all indices are thus equal to $\pm 1$ ($+1$ if $f$ is orientation preserving and $-
1$ if it is orientation reversing at $z_j$). Hence 
\begin{equation}\label{46}
N=n_1+n_{-1}\,,\,\,\,1=n_1-n_{-1}\,,
\end{equation}
implying that
\begin{equation}\label{47}
N=1+2n_{-1}
\end{equation}
is odd.

\subsection{Relations between mean Convergence and reduced Shear}

The Beltrami parameter (reduced shear) $\nu_{f}$ of a lens map is in principle observable. 
What we are really interested in is, however, the mean curvature $\kappa$ which is related to 
the surface mass density by (\protect\ref{210}).

In view of (\protect\ref{318}) it is natural to look first for relations between the complex 
shear $\gamma$ and $\kappa$.

Eq. (\protect\ref{319}) for the differential of the complex lens map and 
(\protect\ref{315}) give
\begin{equation}\label{48}
\gamma = -\bar\partial f = 2\bar\partial^2\psi\,.
\end{equation}
In order to get a useful relation we differentiate (\protect\ref{48}) and use 
(\protect\ref{317})
\begin{equation}\label{49}
\partial\gamma = 2\bar\partial\left(\partial\bar\partial\psi\right) = \bar\partial\kappa\,.
\end{equation}
This can be regarded as an inhomogeneous Cauchy-Riemann equation for
$\kappa$. With the
results in Section 3.1 we conclude
$$
\kappa = \frac{1}{\pi}\left(\frac{1}{z}\right) 
*\partial_{\gamma}+\kappa_{\circ}=\frac{1}{\pi}\partial\left(\frac{1}{z}\right)  
*\gamma+\kappa_{\circ}
$$
or
\begin{equation}\label{410}
\kappa = -\frac{1}{\pi}\frac{1}{z^2}*\gamma+\kappa_{\circ}\,.
\end{equation}
The additive constant $\kappa_{\circ}$ reflects the fact that a homogeneous mass
sheet does 
not produce any shear (`mass sheet degeneracy'). The real form of (\protect\ref{410}) 
appears the first time in \cite{4}. In making use of (\protect\ref{320}), we obtain an integral 
equation for $\kappa$ when $\nu$ is known:
\begin{equation}\label{411}
\kappa = -\frac{1}{\pi}\frac{1}{z^2}*\left\lbrack\nu\left(1-
\kappa\right)\right\rbrack+\kappa_{\circ}\,.
\end{equation}
This has been used, for instance, in \cite{6} for nonlinear cluster inversions.

We add that (\protect\ref{410}) has an inverse, that also appeared in the influencial paper 
\cite{4} of Kaiser and Squires. From (\protect\ref{48}) and (\protect\ref{213})
we obtain
\begin{equation}\label{412}
\gamma = 4\bar\partial^2 G*\kappa\,.
\end{equation}
Since the fundamental solution $G$ of the two-dimensional Laplace operator is
\begin{equation}\label{413}
G=\frac{1}{2\pi}\ln\vert z\vert = \frac{1}{4\pi}\ln\left( z\zbar\right)
\end{equation}
we find
\begin{equation}\label{414}
\gamma = -\frac{1}{\pi}\frac{1}{z^2}*\kappa\,.
\end{equation}

Note that (\protect\ref{49}) has the real form ($\kappa$ is real)
\begin{equation}\label{415}
\nabla\kappa = \left(
\begin{array}{c}
\partial_1 \gamma_1+\partial_2 \gamma_2 \\
\partial_1 \gamma_2+\partial_2 \gamma_1
\end{array}\right)\,.
\end{equation}

Let us differentiate (\protect\ref{49}) once more
\begin{equation}\label{416}
\partial\bar\partial\kappa = \partial^2\gamma\,,
\end{equation}
giving
\begin{equation}\label{417}
\Delta\kappa = 4\partial^2\left\lbrack\nu(1-\kappa)\right\rbrack\,,
\end{equation}
from where we could again arrive at (\protect\ref{411}). The mass-sheet degeneracy is 
reflected in the following invariance property: Eq. (\protect\ref{417}), for given $\nu$, 
remains invariant under the substitution 
\begin{equation}\label{418}
\kappa \longrightarrow \lambda\kappa+(1-\lambda)\,,
\end{equation}
where $\lambda$ is a real constant \cite{14}.

We can use (\protect\ref{49}) in a different manner. First, we write this equation as
$$
\bar\partial\kappa = \partial\left\lbrack\nu(1-\kappa)\right\rbrack = (1-\kappa)\partial\nu-
\nu\partial\kappa\,.
$$
This becomes simpler in terms of $K:=\ln(1-\kappa)$:
\begin{equation}\label{k419}
\bar\partial K-\nu\partial K=\partial\nu\,.
\end{equation}
To this we add its complex conjugate. Noting that $K$ is real, we obtain again an 
inhomogeneous Cauchy-Riemann equation, this time for $K$:
\begin{equation}\label{420}
\bar\partial K = h(\nu)\,,
\end{equation}
whereby the inhomogeneity
\begin{equation}\label{421}
h(\nu) = (1-\vert\nu\vert^2)^{-1}\left\lbrack\partial\nu+\overline{\nu\partial\nu}\right\rbrack
\end{equation}
is in principal observable.

The real form of this equation was obtained by Kaiser \cite{15} and has often been used in 
the analysis of cluster data. The complex version appears also in \cite{7}.

It should have become clear at this point that the complex formulation is also useful. The 
relations, derived in this subsection, emerge alsmost automatically by just applying $\partial$ 
and $\bar\partial$ to the coefficients of the differential of the lens map.

\subsection{Other useful Reconstruction Equations}

Real lensing data are always confined to a finite field of the sky. Therefore, the solution of 
(\protect\ref{420}) in the form (\protect\ref{311}), for example, involving an integration 
over all of $\CC$, is in practice not very useful. One can, however, also obtain integral 
formulas in which only integrations over bounded domains occur.

In order to arrive at these, we write the inhomogeneous Cauchy-Riemann equation in terms of 
differential forms:
\begin{equation}\label{422}
d''g=\omega\,.
\end{equation}
Here $\omega$ is a $1-$form and we use the standard decomposition $d=d'+d''$ of the 
exterior derivative, satisfying
\begin{equation}\label{423}
d'\circ d' =0\,,\,\,\,d''\circ d'' = 0\,,\,\,\,d'\circ d''+d''\circ d' =0
\end{equation}
(see, e.g., \cite{11}). We make also use of the $\star$-operator, which is related to complex 
conjugations as follows: If a $1-$form $\alpha$ is decomposed as 
$\alpha=\alpha_1+\alpha_2$, where $\alpha_1$ is of type $(1,\,0)$ and $\alpha_2$ of type 
$(0,\,1)$, then
\begin{equation}\label{424}
\star\alpha = i(\bar\alpha_1-\bar\alpha_2)\,.
\end{equation}
The following identities are useful:
$$
\star\star\alpha = -\alpha\,,\,\,\,\overline{\star\alpha} = \star\bar\alpha\,,
$$
$$
d\star\left(\alpha_1+\alpha_2\right) = id'\bar\alpha_1-id''\bar\alpha_2\,,
$$
$$
\star d'g = id''\bar g\,,\,\,\,\star d''g = -id'\bar g\,,
$$
\begin{equation}\label{425}
d\star dg = 2id'd''\bar g = \Delta g dx\wedge dy\,,
\end{equation}
where $g$ is a function.

Let now $\Omega \subset \CC$ be a bounded domain with smooth boundary $\partial\Omega$ 
and $A=\vert\Omega\vert$. We show that $g$ minus its average $\bar g$ over $\Omega$,
\begin{equation}\label{426}
\bar g = \frac{1}{A}\int_{\Omega}g dx\wedge dy
\end{equation}
can be represented in the following form
\begin{equation}\label{427}
g-\bar g =\int_{\Omega}\star\alpha\wedge\omega\,.
\end{equation}
The $1-$form $\alpha$ in the integral is given by
\begin{equation}\label{428}
\alpha=2d''H
\end{equation}
in terms of the real Green's function $H$, defined by
\begin{equation}\label{429}
\Delta H-\frac{1}{A}=-\delta\,,
\end{equation}
together with the Neumann boundary condition on $\partial\Omega$. 

This is a consequence of Stokes' theorem. The integrand in (\protect\ref{427}) is
$$
\star\alpha\wedge\omega = \star\alpha\wedge d''g = -d''\left(g\star\alpha\right)+2f 
d''\left(\star d''H\right)\,.
$$
By making use of (\protect\ref{425}) we obtain for the last term
$$
gd''\left(\star d''H\right) = 2ig d''d'H = -g\Delta H dx\wedge dy\,,
$$
while the first term is given by
$$
-d''\left( g\star d''H\right) = -d\left( g\star d''H\right)\,.
$$
Hence,
$$
\int\star\alpha\wedge\omega =-\int_{\partial\Omega}g\star d''H+g-\bar g\,.
$$
This is just (\protect\ref{427}) since the last integral vanishes, due to the Neumann boundary 
condition for $H$. Formulas equivalent to (\protect\ref{427}) have been much used by 
S.~Seitz and P.~Schneider \cite{6}.

The starting point for the derivation of another useful relation is (\protect\ref{317}) in the 
form
$$
d\left( f-z\right) = -\kappa dz-\gamma d\zbar\,,
$$
If we wedge this with $d\zbar$ and add the complex conjugate of the resulting equation we 
find
\begin{equation}\label{430}
\kappa dz\wedge d\zbar = \frac{1}{2}d\left\lbrack\kappa\left( zd\zbar-\zbar 
dz\right)+\gamma \zbar d\zbar-\bar\gamma zdz\right\rbrack\,.
\end{equation}

Taking the average according to (\protect\ref{426}) we arrive at
\begin{equation}\label{431}
\bar\kappa =\left\langle\kappa\right\rangle+\frac{\oint\left(\gamma\zbar d\zbar-\bar\gamma 
zdz\right)}{\oint\left(zd\zbar-\zbar dz\right)}\,,
\end{equation}
where $\left\langle \cdot\right\rangle$ denotes the average along the boundary 
$\partial\Omega$:
\begin{equation}\label{432}
\left\langle\kappa\right\rangle = \frac{\oint\kappa\left(zd\zbar-\bar z 
dz\right)}{\oint\left(zd\zbar-\zbar dz\right)}\,.
\end{equation}
For the special case of a disk $D_r$ we have along the boundary $z=re^{i\varphi}$, 
$zd\zbar-\zbar dz = -2ir^2 d\varphi$, hence
\begin{equation}\label{433}
\bar\kappa = \left\langle\kappa\right\rangle+\left\langle\gamma_t\right\rangle\,,
\end{equation}
where $\gamma_{t}$ denotes the tangential component of the shear
\begin{equation}\label{434}
\gamma_t = \gamma_1\cos 2\varphi+\gamma_2\sin 2\varphi\,.
\end{equation}
This relation is not new (see Ref. \cite{5}). Noting that 
\begin{equation}\label{435}
\bar\kappa =\frac{1}{\pi r^2}\int_{0}^{r}\kappa\left( r',\,\varphi\right) r'dr'd\varphi\,,
\end{equation}
and thus
\begin{equation}\label{436}
\frac{d\bar\kappa}{d\ln r} = 2\left\langle\kappa\right\rangle -2\bar\kappa\,,
\end{equation}
we can use (\protect\ref{433}) to obtain the interesting connection
\begin{equation}\label{437}
\frac{d\bar\kappa}{d\ln r} =-2\left\langle\gamma_t\right\rangle\,.
\end{equation}
This has recently been used in an analysis of weak lensing data \cite{5}. A useful integral 
form of it is, in obvious notation,
\begin{equation}\label{438}
\bar\kappa\left( r_1\right)-\bar\kappa\left( r_1<r<r_2\right) = 2\left(1-
\frac{r_1^2}{r_2^2}\right)^{-
1}\int_{r_1}^{r_2}\left\langle\gamma_t\right\rangle\frac{dr}{r}\,.
\end{equation}
The left hand side of this equation is what Kaiser and Squires call the $\zeta$-statistics, 
$\zeta\left(r_1,\,r_2\right)$. One can use general weight functions for the average process 
\cite{5} and try to optimize the choice for the detection of mass overdensities \cite{6}. Note 
also, that the integral on the right in (\protect\ref{438}) can be written as
\begin{equation}\label{439}
\int_{r_1}^{r_2}\left\langle\gamma_t\right\rangle\frac{dr}{r}=\frac{1}{2\pi}
\int_{\left\lbrack r_1,\,r_2\right\rbrack}\Re\left(\frac{1}{\bar z^2}\bar\gamma\right) 
dx\wedge dy\,.
\end{equation}

We conclude by pointing out another appearance of a Beltrami parameter in lensing theory. 
An often used method for describing the shape of a galaxy image uses the second brightness 
moments
\begin{equation}\label{440}
Q_{ij}=\frac{1}{Norm}\int I\left(\bmx\right)\left( x_i-\bar x_i\right) \left( x_j-\bar 
x_j\right) d^2x\,,
\end{equation}
where $I\left(\bmx\right)$ is the surface brightness distribution and 
$\bar{\bmx}$ is the center of light of the galaxy image. Regard now $Q=\left(Q_{ij}\right)$ as a linear 
map of $\RR^2$. If this is interpreted as a map $z\longmapsto w\left( z\right)$ of $\CC$ it 
reads
\begin{equation}\label{441}
w=\frac{1}{2}\left(Q_{11}+Q_{22}\right) z +\frac{1}{2}\left(Q_{11}-
Q_{22}+2iQ_{12}\right)\zbar = \frac{1}{2}tr Q \left\lbrack z+\chi\zbar\right\rbrack\,,
\end{equation}
where
\begin{equation}\label{442}
\chi = \frac{\left(Q_{11}-Q_{22}+2iQ_{12}\right)}{tr Q}\,.
\end{equation}
$\chi$ is called the complex ellipticity and is clearly just the Beltrami parameter of the map 
(\protect\ref{441}). The intrinsic brightness moments $Q_{ij}^{\left( s\right)}$ of the galaxy 
are defined corespondingly and it is easy to see that $Q^{\left( s\right)} = 
D\varphi\cdot Q\cdot D\varphi$, $D\varphi$ being the differential (\protect\ref{215}) of the 
lens map. The interpretation of $\chi$ just given, allows us to find easily the corresponding 
relation between $\chi$ and $\chi^{\left( s\right)}$. One just has to compose the map 
(\protect\ref{441}) on the right and on the left with the linearized lens map
\begin{equation}\label{443}
w = \left(1-\kappa\right) z-\gamma\zbar\,.
\end{equation}
This gives readily
\begin{equation}\label{444}
\chi^{\left( s\right)}=\frac{-2\nu+\chi+\nu^2\bar\chi}{1+\vert \nu\vert^2-
2\Re\left(\nu\bar\chi\right)}\,,
\end{equation}
with the inverse
\begin{equation}\label{445}
\chi=\frac{2\nu+\chi^{\left( s\right)}+\nu^2\bar\chi^{\left( s\right)}}{1+\vert 
\nu\vert^2+2\Re\left(\nu\bar\chi^{\left( s\right)}\right)}\,.
\end{equation}
A real derivation of these formulas is quite akward. They are used in applications by 
averaging over a set of galaxy images, together with statistical assumptions about the 
intrinsic ellipticity distribution (for instance $\left\langle\chi^{\left( s\right)}\right\rangle 
=0$), to determine the reduced shear $\nu$ of the lens map. Here, we just wanted to point out 
that $\chi$ has the interpretation of a Beltrami parameter and that the relations 
(\protect\ref{444}) and (\protect\ref{445}) are very easily obtained in the complex 
formalism.

We hope that the reader will find other examples of such simplifications.

\section{Acknowledgments}
I thank Philippe Jetzer for interesting conversations and Markus Heusler
for a careful reading of the manuscript. Marcus Str\"assle helped me in
shaping the final version.


\begin{references}
\def\cmp#1#2#3{{ Commun. Math. Phys.} {\bf #1}, #2 (#3)} 
\def\grg#1#2#3{{ Gen. Rel. Grav.} {\bf #1}, #2 (#3)} 
\def\pr#1#2#3{{ Phys. Rev.} {\bf #1}, #2 (#3)} 
\def\prl#1#2#3{{ Phys. Rev. Lett.} {\bf #1}, #2 (#3)} 
\def\prd#1#2#3{{ Phys. Rev. D} {\bf #1}, #2 (#3)} 
\def\pl#1#2#3{{ Phys. Lett} {\bf #1}, #2 (#3)} 
\def\pla#1#2#3{{ Phys. Lett. A} {\bf #1}, #2 (#3)} 
\def\plb#1#2#3{{ Phys. Lett. B} {\bf #1}, #2 (#3)} 
\def\prep#1#2#3{{ Phys. Reports} {\bf #1}, #2 (#3)} 
\def\phys#1#2#3{{ Physica} {\bf #1}, #2 (#3)} 
\def\jcp#1#2#3{{ J. Comput. Phys.} {\bf #1}, #2 (#3)} 
\def\jmp#1#2#3{{ J. Math. Phys.} {\bf #1}, #2 (#3)} 
\def\jpm#1#2#3{{ J. Phys. A: Math. Gen.} {\bf #1}, #2 (#3)} 
\def\cpr#1#2#3{{ Computer Phys. Rept.} {\bf #1}, #2 (#3)} 
\def\cqg#1#2#3{{ Class. Quantum Grav.} {\bf #1}, #2 (#3)} 
\def\cma#1#2#3{{ Computers Math. Applic.} {\bf #1}, #2 (#3)} 
\def\mc#1#2#3{{ Math. Compt.} {\bf #1}, #2 (#3)} 
\def\apj#1#2#3{{ Astrophys. J.} {\bf #1}, #2 (#3)} 
\def\apjs#1#2#3{{ Astrophys. J. Suppl.} {\bf #1}, #2 (#3)} 
\def\acta#1#2#3{{ Acta Astronomica} {\bf #1}, #2 (#3)} 
\def\apl#1#2#3{{Ann. Physik. (Leipzig)} {\bf #1}, #2 (#3)} 
\def\sa#1#2#3{{ Sov. Astro.} {\bf #1}, #2 (#3)} 
\def\sia#1#2#3{{ SIAM J. Sci. Statist. Comput.} {\bf #1}, #2 (#3)} 
\def\aa#1#2#3{{ Astron. Astrophys.} {\bf #1}, #2 (#3)} 
\def\mnras#1#2#3{{ Mon. Not. R. astr. Soc.} {\bf #1}, #2 (#3)} 
\def\npb#1#2#3{{ Nucl. Phys. B} {\bf #1}, #2 (#3)} 
\def\prsla#1#2#3{{ Proc. R. Soc. London, Ser. A} {\bf #1}, #2 (#3)} 
\def\ijmpc#1#2#3{{ I.J.M.P.} C {\bf #1}, #2 (#3)}
%
\def\hepth#1#2{{ hep-th }{\bf #1} (#2)}
\def\grqc#1#2{{ gr-qc }{\bf #1} (#2)}
%
%

\bibitem{1}
P.~Schneider, J.~Ehlers, E.E.~Falco, Gravitational Lenses, Springer-Verlag 
(1992).
\bibitem{2}
P.~Schneider, in: Cosmological Applications of Gravitational Lensing, 
Lecture Notes in Physics, eds. E.~Martinez-Gonz${\rm \acute a}$lez \& J.L.~Sanz, 
Springer-Verlag (1996); P.~Schneider, Helv. Phys. Acta, {\bf 69}, 373
(1996).
\bibitem{3}
N.~Kaiser, Gravitational Lensing, $18^{th}$ Texas Symposium on 
Relativistic Astrophysics, Ann. New York Akad. Sci. (1997), to appear.
\bibitem{4}
N.~Kaiser, G.~Squires, \apj{404}{441}{1993}.
\bibitem{5}
N.~Kaiser, G.~Squires, T.~Broadhurst, \apj{449}{460}{1995}.
\bibitem{6}
S.~Seitz, P.~Schneider, \aa{305}{383}{1996}
\bibitem{7}
T.~Schramm, R.~Kayser, \aa{299}{1}{1995}
\bibitem{8}
N.~Straumann, General Relativity and Relativistic Astrophysics, Springer-
Verlag 1984.
\bibitem{9}
H.~Whitney, Ann. Math., {\bf 62}, 374, (1955). 
\bibitem{10}
L.~Carlson, T.W.~Gamelin, Complex Dynamics, Universitext: Tracts in 
Mathematics, Springer-Verlag (1993).
\bibitem{11}
O.~Forster, Lectures on Riemann Surfaces, Graduate Texts in Mathematics 
81, Springer-Verlag (1981).
\bibitem{12}
P.~Schneider, C.~Seitz, \aa{294}{411}{1995}.
\bibitem{13}
J.B.~Conway, Functions of One Complex Variable, Graduate Texts in 
Mathematics 11, Springer-Verlag (1973).
\bibitem{14}
C.~Seitz, P.~Schneider, \aa{297}{287}{1995}.
\bibitem{15}
N.~Kaiser, \apj{439}{L1}{1995}.

\end{references}
\end{document}